\documentclass[preprint]{vgtc}               

\graphicspath{{figures/}{pictures/}{images/}{./}} 

\usepackage{times}                     

\usepackage{tabu}                      
\usepackage{booktabs}                  
\usepackage{lipsum}                    
\usepackage{mwe}                       

\usepackage{mathptmx}                  

\usepackage{amsmath}

\onlineid{0}

\vgtccategory{Contest}

\vgtcinsertpkg

\preprinttext{Author's version to appear in 2025 IEEE Conference on Virtual Reality and 3D User Interfaces Abstracts and Workshops (VRW).}

\title{Planet Purifiers: A Collaborative Immersive Experience Proposing New Modifications to HOMER and Fishing Reel Interaction Techniques}

\author{
    Alexander Giovannelli\thanks{e-mail: agiovannelli@vt.edu} %
    \and Fionn Murphy
    \and Trey Davis
    \and Chaerin Lee
    \and Rehema Abulikemu
    \and Matthew Gallagher
    \and Sahil Sharma
    \and Lee Lisle
    \and Doug A. Bowman
}
\affiliation{\scriptsize Center for Human-Computer Interaction \\ Department of Computer Science, Virginia Tech, Blacksburg, VA, USA}

\teaser{
  \centering
  \includegraphics[width=\linewidth]{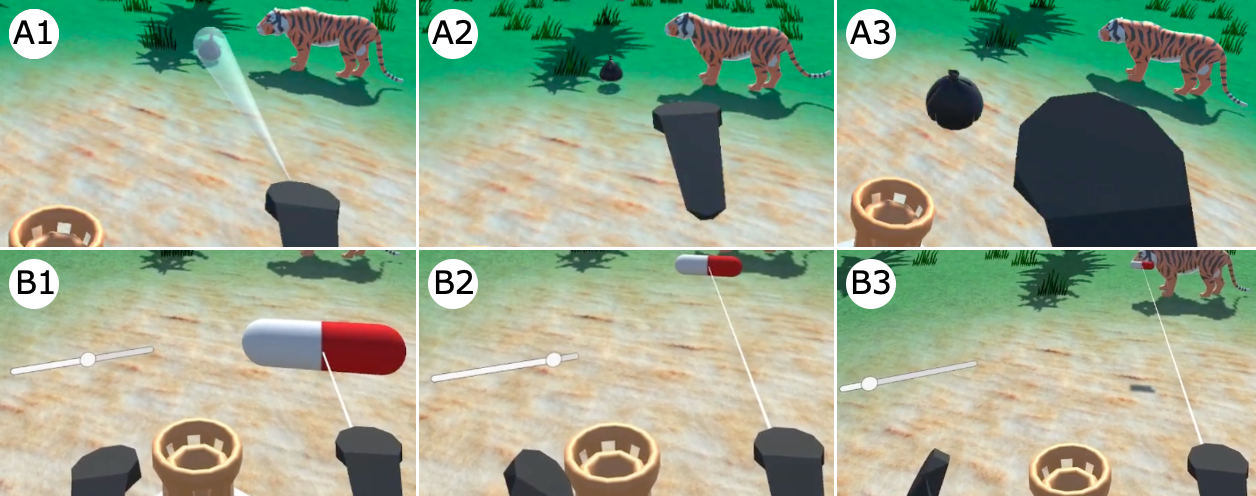}
  \caption{3D interaction techniques proposed in the paper: (A1-A3) FLOW-MATCH, an enhancement to the HOMER technique using a cone volume for initial object selection to improve selection efficiency; (B1-B3) RAWR-XD, an enhancement to the Fishing Reel technique allowing variable translation speed based on the non-selecting wrist tilt for a selected object along a ray-cast.}
  \label{fig:teaser}
}

\abstract{
This paper presents our solution to the 2025 3DUI Contest challenge.
We aimed to develop a collaborative, immersive experience that raises awareness about trash pollution in natural landscapes while enhancing traditional interaction techniques in virtual environments.
To achieve these objectives, we created an engaging multiplayer game where one user collects harmful pollutants while the other user provides medication to impacted wildlife using enhancements to traditional interaction techniques: HOMER and Fishing Reel.
We enhanced HOMER to use a cone volume to reduce the precise aiming required by a selection raycast to provide a more efficient means to collect pollutants at large distances, coined as \textsc{FLOW-MATCH}.
To improve the animal feed distribution to wildlife far away from the user with Fishing Reel, we created \textsc{RAWR-XD}, an asymmetric bi-manual technique to more conveniently adjust the reeling speed using the non-selecting wrist rotation of the user.
} 

\CCScatlist{
  \CCScatTwelve{Human-centered computing}{Human computer interaction (HCI)}{Interaction paradigms}{Virtual reality};
  \CCScatTwelve{Human-centered computing}{Human computer interaction (HCI)}{Interaction techniques}{}
}

\begin{document}



\maketitle

\section{Introduction}
The careless disposal of waste and throwaway culture has significantly impacted native wildlife populations in recent years \cite{tariq2022consequences}.
Although policies have been established to mitigate the severity of the waste produced and to punish improper disposal, concerns on a global scale have only continued to grow.
To highlight the importance of preserving our natural environment and showcase the power of collaboration in building a cleaner Earth, we created an immersive virtual reality experience.
In this gamified experience, two players work together to collect pollutants and heal neighboring wildlife in a future environment devastated by the continued neglect of proper waste disposal practices.
We also propose enhancements to two traditional interaction techniques: HOMER and Fishing Reel.
Our contest entry combines an immersive cooperative experience with efficient and convenient interactions to encourage users to help their surrounding environment in an exciting, easy-to-follow, sci-fi narrative.

\section{Design Process}
To design our contest project, we scheduled two weekly meetings: one focused on group discussions to review project progress and coordinate required deliverables, and the other dedicated to collaborative system development.
Our group discussion meetings entailed using the K-J method to generate affinity diagrams \cite{plain2007build} for brainstorming and consolidating our ideas centered around the contest topic (i.e., promoting environmental sustainability in collaborative virtual environments).
The distillation of the ideas generated from the group discussions then led us to create a backlog of tasks for creating our project.
This backlog was then worked on during our collaboration and development meetings, where we pair-programmed and solicited inter-team feedback on prototyped project increments.

When prototyping, we used Meta Quest head-worn displays and their accompanying Meta Touch Controllers for testing.
We created the final program using the Unity Game Engine \footnote{\url{https://unity.com/}} with the OpenXR framework to allow flexibility in the choice of head-worn display used for the experience.
The networking solution used to support our cooperative experience was Normcore \footnote{\url{https://normcore.io/}}.

\section{Story and Gameplay}
In the year 2077, Earth has been devastated by humanity's failure to stop the improper disposal of toxic materials.
A veil of poisonous smog envelops the natural landscapes, streams of water have become acidic baths, and the wildlife has become rabid and hostile due to mutations from their harsh, waste-ridden surroundings.
Humanity has sought refuge in space vessels while researching how to return to the planet.
Recently, emerging studies have found that removing waste in combination with treating wildlife will allow the planet to be livable once again.
As a result, scientists have worked tirelessly and finally developed a solution: the purifying rover.
This self-piloted vessel is equipped with robotic arms to be controlled by a two-person team: one person to collect the hazardous waste and the other to administer medication to wildlife.

The purifying rover follows a pre-calculated route determined to have a critical impact on the Earth restoration efforts.
It stops at designated cleanup coordinates, opening its protective outer shield to give the two users, acting as scientists, a clear view to perform the cleanup and medicating tasks.
Once the scientists have completed the collection and feeding efforts, or the safe time to have the shield lowered has elapsed, the rover continues until it reaches the end of its route.
At this point, the scientists will be notified of their contributions to the planet purification initiative.

\begin{figure}[h]
    \centering
    \includegraphics[width=\linewidth, alt={A view from above the virtual environment. The landscape is enshrouded in a blanket of orange smog.}]{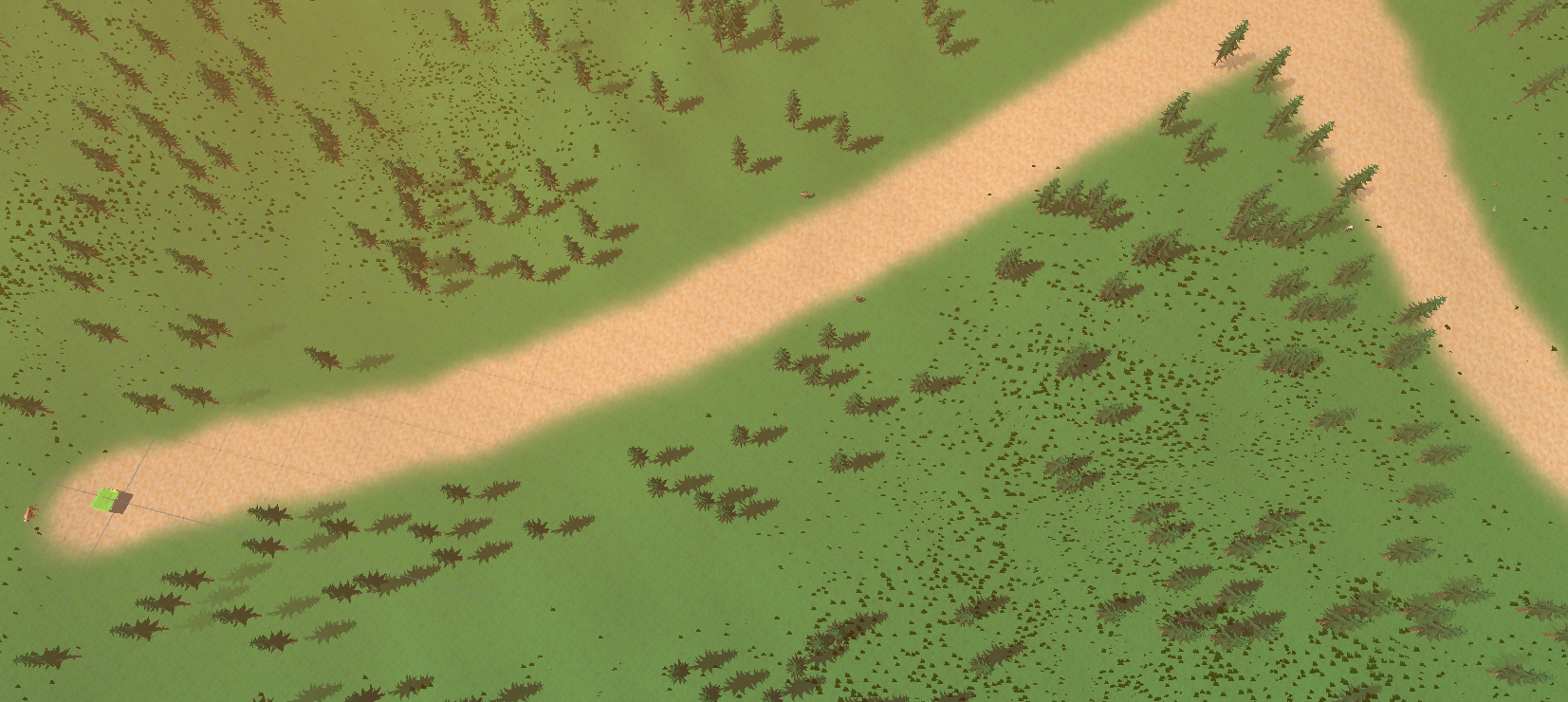}
    \vspace{-5mm}
    \caption{A top-down view of the polluted environment.}
    \vspace{-3mm}
    \label{fig:Environment}
\end{figure}

\section{Interaction Techniques}
The interaction techniques we created for our project were intended to resolve challenges in the efficiency and convenience of target selection and manipulation within 3D virtual environments (VEs).
We focused primarily on (1) the approximate selection and retrieval of distant targets and (2) the distribution of nearby targets to destinations of interest inside VEs.
By referencing existing classical interaction techniques \cite{laviola20173d, May_3DUITK2019}, we found two techniques that could be improved to resolve the aforementioned challenges: Fishing Reel and HOMER \cite{bowman1997evaluation}.

\subsection{FLOW-MATCH}
To provide the pollutant collector experience with the ability to retrieve trash objects in the VE, we first implemented the HOMER technique, a hybrid technique that uses ray-casting for initially selecting a target object and a virtually attached hand to manipulate the transform of the selected object.
The HOMER technique uses the following to manipulate a selected object's position:\\ 
(1) \( SF = \frac{\left\| \overrightarrow{UO} \right\|}{\left\| \overrightarrow{UC} \right\|} \), (2) \( OFF = O_{pos} - \left( U_{pos} + SF \cdot \overrightarrow{UC} \right) \), and (3) \( O_{pos} = U_{pos} + SF \cdot \overrightarrow{UC} + OFF \).\\

The scaling factor (1) is calculated by dividing the user-to-object vector magnitude by the user-to-controller vector magnitude.
The offset (2) is calculated by subtracting the object's initial position from the user's position plus the product of the scale factor and user-to-controller vector.
The object's position (3) is then equal to the user's position plus the product of the scale factor and user-to-controller vector plus the offset.

We found that the initial object selection using HOMER was difficult at large distances due to the high-precision aiming required by the raycast, especially if the target object was small in scale.
To resolve this issue, we replaced the raycast with Flashlight \cite{liang1994jdcad}, a cone volume projected from the controller that requires less precision for the initial selection at long distances.
The combination of the less precise selection afforded by Flashlight and the fluid hand-centered repositioning afforded by HOMER addressed our convenience concerns for trash object selection and manipulation during the experience.
We named this technique ``\textbf{F}lash\textbf{L}ighting \textbf{O}bjects \textbf{W}ith \textbf{MA}nipula\textbf{T}ions \textbf{C}entered at the \textbf{H}and'' or ``\textsc{FLOW-MATCH}''.

\subsection{RAWR-XD}
To distribute medication to afflicted wildlife, we implemented the Fishing Reel technique, which utilizes a virtual ray to select objects and move them back and forth along the ray.
While using this technique, we found that reeling the object along the ray was slow, particularly when moving the object farther away.
To improve the reeling, we extended the technique to be bi-manual, where the wrist's tilt rotation of the non-selecting hand increases or decreases the reeling speed along the ray.
By tilting their wrist left, the base speed at which the object is translated back and forth along the ray will be reduced by a minimum multiplier of 0.5.
By tilting their wrist right, the base speed of object translation along the ray will be multiplied by a maximum of 5.
The multiplier is visually represented by a slider anchored over the top of the user's non-selecting controller.
We named this technique ``\textsc{RAWR-XD}'' or ``\textbf{R}eeling \textbf{A}djustments via the \textbf{W}rist for \textbf{R}ay-casting at e\textbf{X}treme \textbf{D}epth''.

\section{Conclusion}
We created a gamified, immersive VR experience where users collaborate to clean and aid a polluted VE with the intent of promoting environmental consciousness.
Additionally, we created novel enhancements to the HOMER and Fishing Reel interaction techniques to improve efficiency in performing the environment cleanup.


\bibliographystyle{abbrv-doi}

\bibliography{template}
\end{document}